\input{aipcheck}

\documentclass[
    ,final            
  ]
  {aipproc}

\usepackage{epsfig,subfigure}

\layoutstyle{6x9}

\begin{document}

\title{Gauge Messenger Models}


\author{Hyung Do Kim}{
  address={School of Physics and Astronomy and CTP, Seoul National University, Seoul, Korea, 151-747}
}

\begin{abstract}
We consider gauge messenger models in which X and Y gauge bosons and gauginos are messengers of supersymmetry breaking. In simple gauge messenger models, all the soft parameters except $\mu$ and $B\mu$ are calculated in terms of a single scale parameter $M_{\rm SUSY}$ which is proportional to $F/M_{\rm GUT}$. Unique prediction on dark matter in gauge messenger models is discussed. (Based on hep-ph/0601036 and hep-ph/0607169)
 
\end{abstract}

\maketitle


\section{Introduction}

Weak scale supersymmetry is one of the most promising candidates for physics beyond the Standard Model. Gauge coupling is unified in Minimal Supersymmetric Standard Model (MSSM) and radiative electroweak symmetry breaking (EWSB) occurs with large top Yukawa coupling. Furthermore, the lightest supersymmetric particle (LSP) explains the dark matter density of the universe. Undoubtedly MSSM is regarded as the most plausible model completing the Standard Model at the weak scale.

MSSM determines the Higgs potential entirely from soft supersymmetry breaking terms and gauge interactions and it predicts the physical Higgs mass to be lighter than Z boson mass at tree level. Loop induced correction can make the upper bound weaker depending on soft supersymmetry breaking parameters. The LEP bound for the physical Higgs mass is 114.4 GeV and it can be satisfied only with stop masses heavier than 900 GeV or 1 TeV. There appears a discrepancy between the weak scale including the physical Higgs mass ($\sim$ 100 GeV) and soft supersymmetry breaking parameters ($\sim$ 1 TeV). If radiative EWSB is the explanation for the quark and lepton masses, 1 TeV stop drives Higgs mass squared to be too negative. The discrepancy between the physical Higgs mass and large soft supersymmetry breaking parameters ($\sim$ 1 TeV) can be explained only by the cancellation between large supersymmetric mass term $\mu$ for the Higgs ($\sim$ 1 TeV) and large soft supersymmetry breaking parameters ($\sim$ 1 TeV).  Observed weak scale is explained with a degree of fine tuning of a percent. It is the "little hierarchy problem".

The little hierarchy problem can be soften when large mixing between the left and the right-handed stop is considered \cite{Dermisek:2006ey}. The mixing is mainly determined by the ratio of $A_t -\mu \cot \beta$ and stop mass $m_{\tilde{t}}$. For $\tan \beta > 5$, the mixing term is mostly given by the soft tri-linear term $A_t$. The stop mixing provides a finite threshold correction to the quartic coupling of Higgs at the stop mass scale and the physical Higgs mass has a maximum at $|A_t/m_{\tilde{t}}| \sim \sqrt{6}$. With large stop mixing, the LEP bound on the physical Higgs mass can be satisfied even with stop as light as 300 GeV. In the "maximal stop mixing scenario" \cite{Dermisek:2006ey}, the fine tuning in the EWSB is highly reduced. While large stop mixing is good to have a natural EWSB, it is very hard to obtain the large stop mixing at the weak scale unless we provide an extremely large $A_t$ at high energy (several times larger than other soft supersymmetry breaking parameters) or negative stop mass squared $m_{\tilde{t}}^2 < 0$ at the GUT scale. 

The large or maximal stop mixing is possible either by having large $A_t$ at the GUT scale such that even after the exponential damping through the RG running, still the remaining $A_t$ at the weak scale is larger than the stop mass or by lowering the stop mass itself starting from negative stop mass squared. Negative stop mass squared at the GUT scale is driven to be positive by strong interactions if gluino mass is not extremely small. Therefore, the phenomenological consideration guides us into the negative stop mass squared at the GUT scale among vast parameter space.

The most natural parameter space regarding the EWSB lies in the negative mass squared. However, $m^2 < 0$ boundary condition is not compatible with usual assumption of universality in gaugino masses and soft scalar masses. Negative squark mass squared can be quite sizable since it can safely become to be positive with the aid of strong interactions. However, sleptons are different. Especially the right-handed selectron gets a loop correction through the RG running only by U(1) interactions and the contribution is too small.  $m_0^2 \ge -(0.4 M_{1/2})^2$ is the bound at the GUT scale to make all the soft scalar masses to be positive at the weak scale. This problem can be cured if we are away from the universality assumption either in gaugino mass or in soft scalar mass (or both).

Minimal supergravity (mSUGRA) is the framework that has been studied extensively and is regarded as the best benchmark point of the MSSM study due to the minimality of soft parameters coming from the assumption of universality.  Simple assumption is good just for the benchmark scenario but it does not constrain the reality. Confronting the data, it is desirable to explore other possibilities than the most popular benchmark scenario. However, it is very hard to study general parameter space of the MSSM due to too many parameters ($\ge 100$). Instead we can study the detailed prediction on phenomenology if the model is still calculable in terms of a few parameters. Gauge mediation and anomaly mediation were in those classes. All these models have common features that superparticles doing strong interactions are heavier than others.

In this paper we explore a quite different class of models in which all the soft spectrum are more or less degenerate. The gauge messenger model is motivated by i) calculability (or predictability), ii) naturalness, iii) simplicity and iv) a distinctive deviation from mSUGRA.

\section{Gauge messenger Model}

We consider $N=1$ supersymmetric SU(5) grand unified theory  (SUSY GUT) guided by circumstantial evidence of gauge coupling unification \cite{Dermisek:2006qj}. \footnote{Earlier try was given in \cite{Dimopoulos:1982gm}.} There are $N=1$ vector multiplet V transforming as an adjoint of SU(5) and three copies of chiral multiplets $10 +\bar{5}$ and Higgs fields $5_H + \bar{5}_H$. In addition, an adjoint chiral superfield $\Sigma$ is introduced to break $SU(5)$ and supersymmetry. If the adjoint chiral superfield gets its VEV in the scalar component and F component,

\begin{eqnarray}
\Sigma & = & \Sigma_0 
\left( \begin{array}{ccccc}
2 & 0 & 0 & 0 & 0 \\
0 & 2 & 0 & 0 & 0 \\
0 & 0 & 2 & 0 & 0 \\
0 & 0 & 0 & -3 & 0 \\
0 & 0 & 0 & 0 & -3 
\end{array}  \right) 
+ F_{\Sigma} \theta^2
\left( \begin{array}{ccccc}
2 & 0 & 0 & 0 & 0 \\
0 & 2 & 0 & 0 & 0 \\
0 & 0 & 2 & 0 & 0 \\
0 & 0 & 0 & -3 & 0 \\
0 & 0 & 0 & 0 & -3 
\end{array} \right),
\end{eqnarray} 
 
SU(5) gauge symmetry is broken to SU(3)$\times$SU(2)$\times$U(1) and X, Y gauge bosons get their masses at the GUT scale ($g\Sigma_0 \sim M_{\rm GUT}$). In the supersymmetric limit, $\lambda_X$, $\lambda_Y$ gauginos (superpartners of X, Y gauge bosons) also get the same masses. For nonzero F term,  $F_\Sigma \neq 0$, X,Y gauge boson and gaugino masses are split and we get a gauge mediation for the standard model gauge bosons at one loop and sfermions at two loop proportional to $F_{\Sigma}/\Sigma_0$. With an intermediate scale F term ($F_{\Sigma} = (10^{10}{\rm GeV})^2)$, the correct weak scale soft supersymmetry breaking terms are calculated. We define the basic unit of gauge mediation in gauge messenger models as $M_{\rm SUSY}$.

\begin{eqnarray}
M_{\rm SUSY} & = & \frac{\alpha_{\rm GUT}}{4\pi} \left| \frac{F_{\Sigma}}{\Sigma_0} \right| .
\end{eqnarray}

Superparticle masses are calculated using analytic continuation technique into superspace \cite{Giudice:1997ni}. With the minimal setup, the beta function coefficient at the GUT scale is $b_{\rm GUT} = 3$. Gauge mediation is understood as the threshold correction at the messenger scale when we integrate out the messenger field. In gauge messenger models, the messengers (X and Y gauge bosons and gauginos) get their masses at the GUT scale and the messenger scale is the GUT scale. More precisely the messengers are massive X and Y gauge bosons and gauginos which are made of massless X and Y gauge bosons and gauginos and massless $\Sigma$.

Gaugino masses at the GUT scale is simply given by the difference of the beta function coefficient above and below the GUT scale. The change occurs only in V and $\Sigma$ and is calculated to be $\Delta b_i = 10 - 2N_{C_i}$ since $b = 10$ in SU(5) with V and $\Sigma$ and $b=2N_{C_i}$ in SU($N_{C_i}$). Note that the components commuting with $T_{24}$ does not get their masses from $\Sigma$.

\begin{eqnarray}
M_i & = & - \Delta b_i M_{\rm SUSY},
\end{eqnarray}
and more explicitly,
\begin{eqnarray}
M_3 & = & -4 M_{\rm SUSY}, \\
M_2 & = & -6 M_{\rm SUSY}, \\
M_1 & = & -10 M_{\rm SUSY}.
\end{eqnarray}

Soft tri-linear terms are also easily calculated by considering the change of the anomalous dimensions of the three corresponding fields.

\begin{eqnarray}
A_{ijk} & = & A_j + A_j + A_k, \\
A_i & = & 2 \Delta c_i M_{\rm SUSY},
\end{eqnarray}
and more specifically,
\begin{eqnarray}
A_t & = & 10 M_{\rm SUSY}, \\
A_b & = & 8 M_{\rm SUSY}, \\
A_{\tau} & = & 12 M_{\rm SUSY}.
\end{eqnarray}

Note the minus sign in gaugino masses. The contribution of massive vector messengers is the opposite of the one of massive chiral messengers. However, the overall sign can be rotated away by $U(1)_R$ symmetry and $A$ term and $\mu$ term change sign accordingly. Only the relative sign of gauginos and $A$ terms is important. From now on, we take gaugino mass to be positive. Accordingly $A$ terms are negative at the GUT scale.

\begin{eqnarray}
M_3 & = & 4 M_{\rm SUSY}, \\
M_2 & = & 6 M_{\rm SUSY}, \\
M_1 & = & 10 M_{\rm SUSY}, \\
A_t & = & -10 M_{\rm SUSY}, \\
A_b & = & -8 M_{\rm SUSY}, \\
A_{\tau} & = & -12 M_{\rm SUSY}.
\end{eqnarray}

Soft scalar mass squared parameters are also calculated as the threshold correction at the GUT scale but now there are two pieces, one is the change of the beta function and the other is the change of the anomalous dimension. Let me summarize the result here. Detailed derivation and discussion is in \cite{Dermisek:2006qj}.

\begin{eqnarray}
m_Q^2 & = & -11 M_{\rm SUSY}^2, \\
m_{u^c}^2 & = & -4 M_{\rm SUSY}^2, \\
m_{d^c}^2 & = & -6 M_{\rm SUSY}^2, \\
m_L^2 & = & -3 M_{\rm SUSY}^2, \\
m_{e^c}^2 & = & 6 M_{\rm SUSY}^2, \\
m_{H_u,H_d}^2 & = & -3 M_{\rm SUSY}^2.
\end{eqnarray}

\begin{figure}[!t]
  \resizebox{.6\textwidth}{!}
  		  {\includegraphics{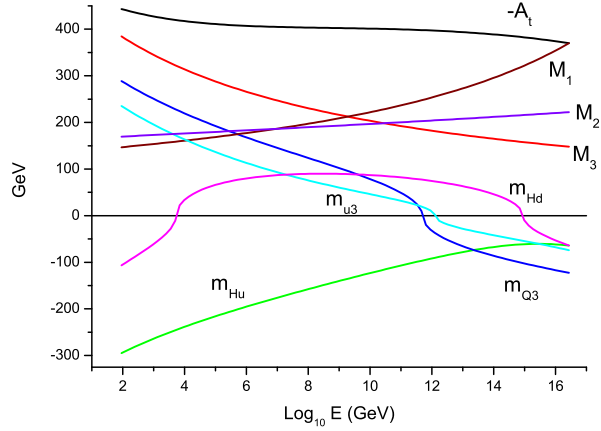}}
 \source{Dermisek,Kim,Kim}
\vspace{-0.5cm}
\caption{ \label{fig:1G_RGhiggs}
Renormalization group running of soft SUSY breaking parameters for simple gauge mediation with
$M_{\rm SUSY} = 37~{\rm GeV}$ and $\tan \beta = 23$. 
Evolution of gaugino masses, $A_t$, and stop and Higgs soft masses are shown here.
In order to have both mass dimension one and two parameters on the same plot
and keep information about signs, we define $m_{H_u} \equiv
m_{H_u}^2/\sqrt{|m_{H_u}^2|} $ and similarly for other scalar masses.
}
\end{figure}

\begin{figure}[!t]
  \resizebox{.6\textwidth}{!}
  		  {\includegraphics{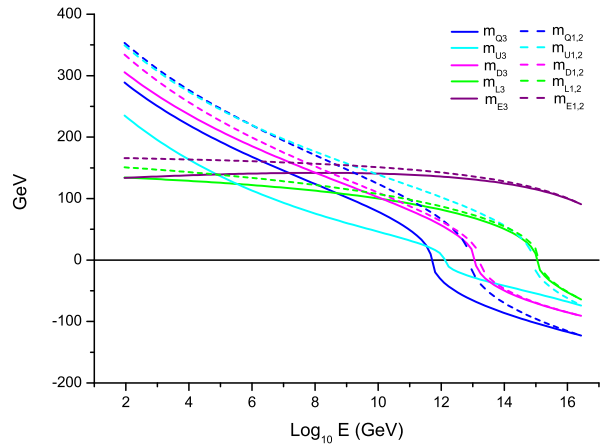}}
		   \source{Dermisek,Kim,Kim}
\vspace{-0.5cm}
\caption{ \label{fig:2G_RGrunning}
The same parameters as in fig. 1 and now evolutions of squark and slepton masses of the third generation (solid lines)
and the first two generations (dashed lines) are shown.}
\end{figure}

Gaugino mass, A term and soft scalar mass squared are calculated and are determined only by a single parameter $M_{\rm SUSY}$ at the messenger scale (the GUT scale). The previous consideration for fine tuning required negative stop mass squared with not so negative slepton mass squared. Indeed the calculated spectrum satisfy the condition. Furthermore, the gaugino masses have the desirable pattern such that heavy bino and wino compared to gluino at the GUT scale can make the left-handed slepton to be positive enough at the weak scale. The characteristic features are summarzied as follows.

\begin{enumerate}
\item Non-universal gaugino masses at the GUT scale : $M_1 : M_2 : M_3 = 5:3:2$.

After the running into the weak scale, the ratio becomes
\begin{eqnarray}
M_1 : M_2 : M_3 & \simeq & 1 : 1.1 : 2.4.
\end{eqnarray}
All the gaugino masses are squeezed (heaviest/lightest = 2.4) compared to other scenarios whether the ratio is in the range of 6 to 10. Thus we expect relatively light gluino with the same constraint on bino or wino. Furthermore, wino/bino mass ratio is 1.1 and these two particles are degenerate within 10 percent. It has a dramatic implication in neutralino dark matter \cite{Bae:2006}.

\item Negative squarks and sleptons masses squared at the GUT scale :

Squarks are more negative since X and Y gauge bosons and gauginos contribute more to them. The pattern results in very squeezed spectrum at the weak scale. The one that gets larger RG running contribution started from more negative value and can be the same as the one that gets smaller contribution. All the squark and slepton masses are lying within factor two at the weak scale. The light stop can be even lighter than sleptons which is impossible in mSUGRA framework.

\item Large A term at the GUT scale :

Unlike the usual gauge mediation where A term is absent at the messenger scale and is generated only through the RG running, large A term is possible in gauge messenger models since the gauge group changes at the messenger scale. It helps to achieve the large (or maximal) stop mixing and can lower the needed stop mass to be lighter but at the same time the large A term itself contributes to the running of up type Higgs soft mass squared and make it to be more negative. Therefore, the fine tuning can be improved only up to certain levels (10 percent in our study).

\end{enumerate}

\subsection{Simple Gauge Messenger Model}

\begin{figure}[!t]
   \resizebox{.6\textwidth}{!}
  		  {\includegraphics{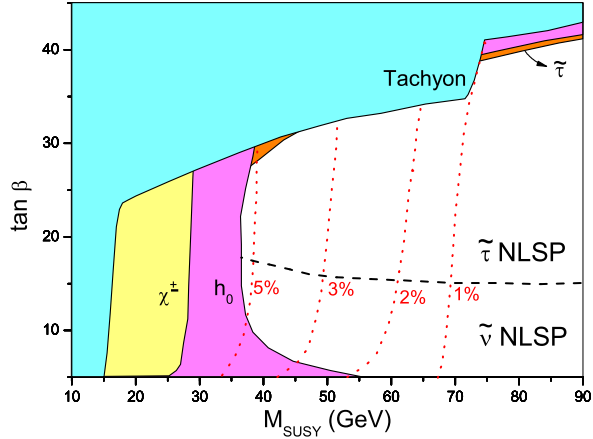}}
		  \source{Dermisek,Kim,Kim}
\vspace{-0.5cm}
\caption{ \label{fig:3Gspace}
Allowed region of parameter space and the degree of fine tuning in
the $M_{\rm SUSY}-\tan \beta$ plane for simple gauge mediation.
The shaded regions are excluded by direct searches for SUSY and Higgs particles.
The region denoted as ``tachyon" is  excluded  due to tachyonic spectrum.
The black dashed line separates regions where sneutrino or stau is NLSP.
}
\end{figure}

Although the gravity mediation is not entirely negligible due to the high messenger scale, first let us deal with the unambiguously calculable parts which we name as a simple gauge messenger model. As we do not address the $\mu$ problem, $\mu$ and $B\mu$ are independent parameters in addition to $M_{\rm SUSY}$. We can determine all the soft spectrum in terms of three parameters. We can trade $B\mu$ with $\tan \beta$ and $\mu$ with $M_Z$. Therefore, with the measured $M_Z$, we have two degrees to vary to explore the parameter space which are $M_{\rm SUSY}$ and $\tan \beta$.

RG running of soft supersymmetry breaking parameters are plotted in figure 1 and 2. Squarks, Sleptons and Higgs start from negative mass sqaured at the GUT scale and end up with positive mass squared except Higgs fields. Only right-handed sleptons start from positive mass squared at the GUT scale and get very small correction through RG running. Detailed spectrum is in the table of \cite{Dermisek:2006qj}.

Figure 3 shows the parameter space in which allowed regions and corresponding next to the lightest particle (NLSP) is given. Gravitino mass is at around a few 10 to 100 GeV and remains as the lightest supersymmetric particle (LSP). Mainly there are four candidates for the NLSP. Neutralino, stau, sneutrino and stop. In simple gauge mediation, the bino/wino mass is slightly heavier than sleptons and also $\mu$ term is larger. Thus neutralino can not be an (N)LSP. Among sleptons, left-handed ones are slightly lighter than the right-handed one. After including D-term contribution, it is the tau sneutrino which becomes an NLSP. For $\tan \beta \le 15$, indeed tau sneutrino is the NLSP. If $\tan \beta \ge 15$, the mixing between the left and the right-handed stau is not negligible and the lightest stau becomes lighter than the tau sneutrino. Therefore, For $\tan \beta \ge 15$, the mixed stau becomes the NLSP. The implication of sneutrino and stau NLSP in gauge messenger models to cosmology and LHC is huge. The NLSP with the weak scale gravitino mass decays very late (life time longer than $10^4$ sec) and might be dangerous for the successful big bang nucleosynthesis (BBN). If the relic density of the NLSP is small enough, the effect on BBN would be harmless. Careful analysis is needed on this topic.

\subsection{Gravity Contribution in the Higgs Sector}

\begin{figure}[!t]
   \resizebox{.6\textwidth}{!}
   		  {\includegraphics{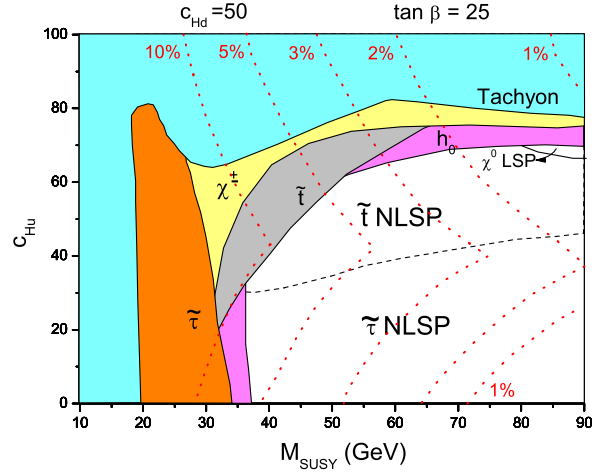}}
 \source{Dermisek,Kim,Kim}
\vspace{-0.5cm}
\caption{ \label{fig:4GHspace}
Allowed region of parameter space and the degree of fine tuning in
the $M_{\rm SUSY}-\tan \beta$ plane for gauge mediation with a gravity mediation to the Higgs sector.
The black dashed line separates regions where sneutrino or stau is NLSP.
}
\end{figure}

Giudice-Masiero mechanism is known as the best solution to the $\mu$ problem and it works only with gravity mediation. Therefore, it is the most desirable to separate matter fields from Higgs such that Higgs fields directly feel the gravity mediation comparable to gauge messenger contribution but matter fields just get their soft masses from gauge messengers. Figure 4 shows the change of the parameter space for fixed $\tan \beta =25$ and $c_{H_d} = 50$. ($\Delta m_{H_d}^2 = c_{H_d} M_{\rm SUSY}^2$) As we chose $\tan \beta \ge 15$, the NLSP is the stau when $c_{H_u}$ is small. When $c_{H_u}$ is larger than 30, the NLSP becomes the stop rather than the stau since top Yukawa driven correction proportional to $m_{H_u}^2$ makes the stop lighter than before. Increasing $c_{H_u}$ also has an impact on $\mu$ since $m_{H_u}^2$ can be small (and negative) by starting from larger value at the GUT scale. There is a tiny region where neutralino is (N)LSP for $M_{\rm SUSY} = 80, 90$ and $c_{H_u} = 60$. The most interesting prediction of the gauge messenger models is the stop NLSP which can not be imagined in the framework of mSUGRA.

\subsection{mSUGRA contribution}

The gravity mediation is not entirely negligible and is expected to modify the prediction of the simple gauge messenger models. If sequestering is possible (separation of matter fields from supersymmetry breaking source), the simple gauge messenger models are realized. Instead if the separation occurs in the Einstein frame, there are universal common $m_{3/2}^2$ (gravitino mass) to all the soft scalar mass squared. Even small contribution changes the parameter space such that neutralino becomes an (N)LSP in most of the parameter space. It is very clear that common universal mSUGRA contribution just raises the mass of scalar particles like sneutrino, stau and stop. The effect on gaugino mass is absent and on $\mu$ is also not significant and neutralino remains light. In this case we can realize the neutralino (N)LSP scenario which is quite different from the neutralino LSP from mSUGRA \cite{Bae:2006}.

\section{Conclusion}

We studied gauge messenger models in which heavy gauge fields are messengers of supersymmetry breaking in SU(5) SUSY GUT models. Larger bino and wino compared gluino mass at the GUT scale is predicted and it ends up with nearly degenerate bino/wino and twice heavier gluino mass at the weak scale. Also it predicts negative squark mass squared and helps to reduce the fine tuning in the electroweak symmetry breaking. The simple predictable model is presented guided by phenomenology and naturalness and it indicates that we might live in a meta-stable vacuum while the deeper minimum exists where color and charge is broken. The model predicts quite a different (N)LSPs in most of the parameter space and the expected signature at LHC is dramatically different from conventional benchmark scenarios based on mSUGRA. We expect that LHC will tell us whether the most natural region of the parameter space pointed out by gauge messenger models is chosen by nature soon.





\bibliographystyle{aipproc}   

\end{document}